\documentclass[english,aps,tightenlines,showpacs, showkeys, notitlepage]{revtex4-1}
\usepackage[latin9]{inputenc}
\usepackage{amssymb}

\makeatletter

\usepackage{dcolumn}
\usepackage{bm}

\makeatother

\usepackage{babel}
\begin{document}
\title{Generalized cosmological constant from gauging Maxwell-conformal algebra}
\author{Salih Kibaro\u{g}lu}
\email{salihkibaroglu@gmail.com}

\author{Oktay Cebecio\u{g}lu}
\email{ocebecioglu@kocaeli.edu.tr}

\date{\today}
\begin{abstract}
The Maxwell extension of the conformal algebra is presented. With
the help of gauging the Maxwell-conformal group, a conformally invariant
theory of gravity is constructed. In contrast to the conventional
conformally invariant actions, our gravitational action contains the
Einstein-Hilbert term without introducing any additional (compensator)
scalar field to satisfy the local scale invariance. This is achieved
by using the curvatures of the algebra. In a special condition, we
show that the resulting action is reduced to the Brans-Dicke like
theory of gravity. We subsequently find the generalized Einstein field
equation together with a coordinate dependent cosmological term and
additional contributions.
\end{abstract}
\affiliation{Department of Physics, Kocaeli University, 41380 Kocaeli, Turkey}
\maketitle

\section{Introduction}

One of the essential problems of modern high-energy physics is to
construct a consistent theory which contains gravity together with
other interactions in a single framework. To deal with this challenge,
one needs to propose a unified theory which governs the gravitational
and quantum behaviours of nature. In this context, we have two different
theories named as General Theory of Relativity (GR) and Quantum Field
Theory (QFT). Both of them have been very successful in their conceptual
frameworks with several shortcomings in extreme regimes. To unify
these theories, we need to reformulate or modify them because there
are some difficulties (for more detail see \citep{capozziello2011}).
For this purpose, the symmetry feature of nature may play a significant
role in the mentioned problem because symmetries simplify the mathematical
complexity of the physical systems \citep{wess1960}. One of the best
examples is the standard model in which local symmetry plays the predominant
role. All of the four fundamental interactions in nature, except gravity,
are described by this model. This theory based on the global gauge
symmetry principle. Furthermore, motivating from Yang and Mills \citep{yang1954},
understanding the gravitational interaction as a gauge theory started
with Utiyama's work \citep{utiyama1956}. He proposed that GR can
be obtained as a gauge theory based on the local homogeneous Lorentz
group. From this idea, Kibble and Sciama constructed a gauge theory
of gravity which contains GR with non-zero torsion based on localization
of the Poincare group \citep{kibble1961,sciama1964}. After these
significant works, many space-time groups were analyzed in the gauge
theoretical context such as the Weyl \citep{Charap1974,Freund1974,bramson1974,omote1974,kasuya1975,Bicknell1976},
affine \citep{borisov1974,hehl1978,lord1978} and conformal \citep{Crispim-Romao1977,kaku1977,kaku1977b,kaku1978,marchildon1978,mansouri1979,nieh1982A,lord1985,wheeler1991,kerrick1995}
groups. These studies lead to gravitational theories more general
than GR. This idea is important because considering gravity as a theory
based on local space-time symmetries corresponds to a considerable
advance towards the unified theory \citep{Blagojevic2002}.

It is known that the conformal transformations play an important role
in physics. For instance, in order to deal with the unification problem,
the conformal gauge theory of gravity is seen a good candidate \citep{flato1970,Alfaro1980,obukhov1982,frankin1985,shtanov1994,hooft2011}
since it is expected to be renormalizable. Furthermore, there are
several uses of the conformal group $\mathcal{C}\left(1,3\right)$
such as the quantum field theory in the high energy regime, cosmological
constant problem \citep{shtanov1994}, dark matter and dark energy
\citep{mannheim2006,mannheim2012}. For these reasons, $\mathcal{C}\left(1,3\right)$
has reached much interest and extensive study in the literature. In
the gravitational sector, the conformal gravity can be seen as a natural
extension of GR with additional symmetries \citep{bekenstein1980,padmanabhan1985,kazanas1991,maldacena2011}
but it has some restrictions for the construction an invariant action.
Invariance of the gravitational action for the local conformal transformations
can be constructed by the terms either linear or quadratic in curvature
forms. In curvature-linear actions, we need to introduce a Brans-Dicke
like compensating scalar field to satisfy the scale invariance \citep{brans1961,dirac1973}.
This approach leads to generalizing Einstein's field equations and
it can be reduced to the standard GR in a special condition. On the
other side, if we use a curvature-quadratic action that can be constructed
by the square of the Weyl tensor, we get higher-order theory of gravity
which does not contain Einstein's gravity (more detail see \citep{wehner1999}). 

To avoid the mentioned restrictions on GR, we consider extending the
conformal group with the Maxwell symmetry. The Maxwell group $\mathcal{M}\left(1,3\right)$
is a symmetry extension which contains non-commutative momentum operator
as $\left[P_{a},P_{b}\right]=iZ_{ab}$ \citep{soroka2005,gomis2009,bonanos2009,bonanos2010,bonanos2010A}.
Here, the antisymmetric central charges $Z_{ab}$ represent additional
degrees of freedom. According to the early studies \citep{bacry1970,schrader1972},
the Maxwell symmetry has been assumed to describe a particle moving
in a Minkowski space-time filled with a constant electromagnetic background
field. In recent years the Maxwell group and its modifications have
gained much interest in the context of producing an alternative way
to generalize Einstein's theory of gravity and supergravity \citep{azcarraga2011,soroka2012,durka2012,Azcarraga2014,cebecioglu2014,salgado2014,cebecioglu2015,concha2015A,penafiel2018,ravera2018,concha2019A,concha2019B,kibaroglu2019A,kibaroglu2019B}.
In particular, gauging the Maxwell (super)algebra provide an effective
background to obtain generalized cosmological constant term \citep{azcarraga2011,soroka2012,durka2012,cebecioglu2014,concha2015A,penafiel2018,kibaroglu2019B}.
In this extended framework, the new degrees of freedom represent uniform
gauge field strengths in (super)space which leading to uniform constant
energy density \citep{bonanos2010}. Therefore, we can say that the
Maxwell symmetry allows a powerful framework to generalize GR. Besides,
this symmetry can be used in various areas such as describing planar
dynamics of the Landau problem \citep{fedoruk2012}, the higher spin
fields \citep{fedoruk2013,fedoruk2013A}, and it was also used in
the string theory as an internal symmetry of the matter gauge fields
\citep{hoseinzadeh2015}.

Our main motivation on this extension is to develop a theory respecting
the local conformal symmetry. For this, we imposed certain constraints
in order to break the initial symmetry. After the symmetry breaking,
the action in the present work yields the generalized Brans-Dicke
theory when the Maxwell-conformal\textbf{$\mathcal{MC}\left(1,3\right)$}
symmetry is broken down to the conformal group. 

The organization of this letter is as follows. In Section II, we briefly
recall the conformal algebra $\mathfrak{c}\left(1,3\right)$ and we
then obtain the Maxwell-Conformal algebra $\mathfrak{mc}\left(1,3\right)$
by taking account of the isomorphism between $\mathcal{C}\left(1,3\right)$
and $\mathcal{SO}\left(2,4\right)$ groups. In Section III, the gauge
theory of $\mathcal{MC}\left(1,3\right)$ group is constructed. In
Section IV, after a brief discussion of the conformal gravity, we
establish a new conformal invariant action that contains the Einstein-Hilbert
term together with the contributions which come from the Maxwell symmetry.
We show that the action can be reduced to a Brans-Dicke like theory
of gravity by considering some constraints. We also give a generalized
version of the Einstein field equations together with a dynamical
cosmological term and an additional energy-momentum tensors. In the
last section, we conclude our work with some comments and possible
future developments.

\section{The conformal algebra and its maxwell extension}

In this section, we consider the conformal group $\mathcal{C}\left(1,3\right)$
and its Maxwell extension in four-dimensional space-time. At first,
we give a brief summary of $\mathcal{C}\left(1,3\right)$. The conformal
group was introduced into physics by Bateman and Cunningham \citep{cunningham1910,bateman1910A,bateman1910B}
in the framework of the electromagnetic theory. It can also be interpreted
as the largest group of coordinate transformation which preserves
the light-cone structure \citep{O'Hanlon1973}. This symmetry contains
the Lorentz, translations, special conformal transformations, and
dilation transformations. Furthermore, the Lie groups of $\mathcal{C}\left(1,3\right)$
and $\mathcal{SO}\left(2,4\right)$ are locally isomorphic, so they
satisfy the same algebra. Since the generators $J_{AB}$ of $\mathcal{SO}\left(2,4\right)$
group satisfy the following commutation relations,

\begin{equation}
\left[J_{AB},J_{CD}\right]=i\left(\eta_{AD}J_{BC}+\eta_{BC}J_{AD}-\eta_{AC}J_{BD}-\eta_{BD}J_{AC}\right),\label{eq: commutation J_AB}
\end{equation}
where the metric $\eta_{AB}=diag\left(+,-,-,-,-,+\right)$ and the
capital Latin indices run $A,B,...=0,1,2,3,5,6$. The generators of
conformal transformations are represented by $X_{A}=\left\{ P_{a},K_{a},D,M_{ab}\right\} $
which correspond to the translations, special conformal transformations
dilatations and Lorentz transformations can be realized through $J_{AB}$
generators as follows \citep{srivastava1973,furlan1979}, 

\begin{eqnarray}
M_{ab} & = & J_{ab},\,\,\,\,\,\,\,\,D=J_{65},\nonumber \\
P_{a} & = & \frac{1}{l}\left(J_{5a}+J_{6a}\right),\,\,\,\,\,\,\,\,K_{a}=l\left(-J_{5a}+J_{6a}\right),\label{eq: conf gen}
\end{eqnarray}
where $l$ is the unit of length ($L$). Thus, the commutation relations
for the generators in Eq.(\ref{eq: conf gen}) can be written as follows,

\begin{eqnarray}
\left[M_{ab},M_{cd}\right] & = & i\left(\eta_{ad}M_{bc}+\eta_{bc}M_{ad}-\eta_{ac}M_{bd}-\eta_{bd}M_{ac}\right),\nonumber \\
\left[M_{ab},P_{c}\right] & = & i\left(\eta_{bc}P_{a}-\eta_{ac}P_{b}\right),\nonumber \\
\left[M_{ab},K_{c}\right] & = & i\left(\eta_{bc}K_{a}-\eta_{ac}K_{b}\right),\nonumber \\
\left[P_{a},P_{b}\right] & = & 0,\,\,\,\,\,\,\left[K_{a},K_{b}\right]=0,\nonumber \\
\left[P_{a},K_{b}\right] & = & 2i\left(\eta_{ab}D-M_{ab}\right),\nonumber \\
\left[M_{ab},D\right] & = & 0,\,\,\,\:\,\left[D,D\right]=0,\nonumber \\
\left[P_{a},D\right] & = & iP_{a},\,\,\,\,\,\left[K_{a},D\right]=-iK_{a}.
\end{eqnarray}
Therefore, we get the usual conformal Lie algebra $\mathfrak{c}\left(1,3\right)$.
Here, $\eta_{ab}=diag\left(+,-,-,-\right)$ and the small Latin indices
have the range $a,b,...=0,1,2,3$. Analogously to the approach taken
above, one can extend $\mathcal{SO}\left(2,4\right)$ algebra in Eq.(\ref{eq: commutation J_AB})
by an additional antisymmetric central charges $\mathcal{Z}_{AB}$
as follows \citep{soroka2012,salgado2014},

\begin{eqnarray}
\left[\mathcal{J}_{AB},\mathcal{J}_{CD}\right] & = & i\left(\eta_{AD}\mathcal{J}_{BC}+\eta_{BC}\mathcal{J}_{AD}-\eta_{AC}\mathcal{J}_{BD}-\eta_{BD}\mathcal{J}_{AC}\right),\nonumber \\
\left[\mathcal{J}_{AB},\mathcal{Z}_{CD}\right] & = & i\left(\eta_{AD}\mathcal{Z}_{BC}+\eta_{BC}\mathcal{Z}_{AD}-\eta_{AC}\mathcal{Z}_{BD}-\eta_{BD}\mathcal{Z}_{AC}\right),\nonumber \\
\left[\mathcal{Z}_{AB},\mathcal{Z}_{CD}\right] & = & 0,\label{eq: MSO(2,4)}
\end{eqnarray}
where $\mathcal{J}_{AB}$ correspond to a generalized $\mathcal{SO}\left(2,4\right)$
generators. The resulting extended algebra can be named as the Maxwell
extension of $\mathfrak{so}\left(2,4\right)$ algebra ($\mathfrak{mso}\left(2,4\right)$).
In order to find the Maxwell extension of $\mathfrak{c}\left(1,3\right)$
algebra, decomposing $\mathcal{J}_{AB}$ and $\mathcal{Z}_{AB}$ into
the following generators,

\begin{eqnarray}
M_{ab} & = & \mathcal{J}_{ab},\,\,\,\,\,\,\,\,D=\mathcal{J}_{65},\nonumber \\
P_{a} & = & \sqrt{\lambda}\left(\mathcal{J}_{5a}+\mathcal{J}_{6a}+\frac{1}{2}\mathcal{Z}_{5a}\right),\nonumber \\
K_{a} & = & \frac{1}{\sqrt{\lambda}}\left(-\mathcal{J}_{5a}+\mathcal{J}_{6a}-\frac{1}{2}\mathcal{Z}_{5a}\right),\nonumber \\
Z_{ab} & = & \mathcal{Z}_{ab},\,\,\,\,\,\,\,\,\,Z=\mathcal{Z}_{65},\nonumber \\
Z_{a} & = & \sqrt{\lambda}\left(\mathcal{Z}_{5a}+\mathcal{Z}_{6a}\right),\nonumber \\
\tilde{Z}_{a} & = & \frac{1}{\sqrt{\lambda}}\left(-\mathcal{Z}_{5a}+\mathcal{Z}_{6a}\right),
\end{eqnarray}
and checking all possible commutators of the generators, the Lie algebra
of the corresponding group is found to be

\begin{eqnarray}
\left[M_{ab},M_{cd}\right] & = & i\left(\eta_{ad}M_{bc}+\eta_{bc}M_{ad}-\eta_{ac}M_{bd}-\eta_{bd}M_{ac}\right),\nonumber \\
\left[M_{ab},P_{c}\right] & = & i\left(\eta_{bc}P_{a}-\eta_{ac}P_{b}\right),\nonumber \\
\left[M_{ab},K_{c}\right] & = & i\left(\eta_{bc}K_{a}-\eta_{ac}K_{b}\right),\nonumber \\
\left[P_{a},P_{b}\right] & = & i\lambda Z_{ab},\,\,\,\,\,\,\left[K_{a},K_{b}\right]=\frac{i}{\lambda}Z_{ab},\nonumber \\
\left[P_{a},K_{b}\right] & = & 2i\left(\eta_{ab}D-M_{ab}\right)+i\left(\eta_{ab}Z-Z_{ab}\right),\nonumber \\
\left[P_{a},D\right] & = & i\left(P_{a}+\frac{\lambda}{2}\tilde{Z}_{a}\right),\,\,\,\,\,\,\left[K_{a},D\right]=-i\left(K_{a}+\frac{1}{2\lambda}Z_{a}\right),\nonumber \\
\left[Z_{a},D\right] & = & iZ_{a},\,\,\,\,\,\left[\tilde{Z}_{a},D\right]=-i\tilde{Z}_{a},\nonumber \\
\left[M_{ab},Z_{cd}\right] & = & i\left(\eta_{ad}Z_{bc}+\eta_{bc}Z_{ad}-\eta_{ac}Z_{bd}-\eta_{bd}Z_{ac}\right),\nonumber \\
\left[M_{ab},Z_{c}\right] & = & i\left(\eta_{bc}Z_{a}-\eta_{ac}Z_{b}\right),\nonumber \\
\left[M_{ab},\tilde{Z}_{c}\right] & = & i\left(\eta_{bc}\tilde{Z}_{a}-\eta_{ac}\tilde{Z}_{b}\right),\nonumber \\
\left[Z_{ab},P_{c}\right] & = & i\left(\eta_{bc}Z_{a}-\eta_{ac}Z_{b}\right),\nonumber \\
\left[Z_{ab},K_{c}\right] & = & i\left(\eta_{bc}\tilde{Z}_{a}-\eta_{ac}\tilde{Z}_{b}\right),\nonumber \\
\left[P_{a},\tilde{Z}_{b}\right] & = & 2i\left(\eta_{ab}Z-Z_{ab}\right),\nonumber \\
\left[K_{a},Z_{b}\right] & = & -2i\left(\eta_{ab}Z+Z_{ab}\right),\nonumber \\
\left[P_{a},Z\right] & = & iZ_{a},\,\,\,\,\,\left[K_{a},Z\right]=-i\tilde{Z}_{a},\label{eq: mc algebra}
\end{eqnarray}
while the remaining commutators are zero. Here, the constant $\lambda$
have the unit of $L^{-2}$ which will be considered as the cosmological
constant. This algebra can be called as the Maxwell-conformal algebra
$\mathfrak{mc}\left(1,3\right)$. Here, the additional generators
$Z_{ab}$, $Z_{a}$, $\tilde{Z}_{a}$ and $Z$ represent tensor, vector
and scalar contributions of the Maxwell symmetry, respectively. The
self-consistency of this algebra can be checked by the help of the
Jacobi identities. 

We also note that the decomposition of the higher dimensional tensor
extended Maxwell algebras provide a powerful framework to generate
various forms of the Maxwell algebras. For example, if we choose the
generators as

\begin{eqnarray}
M_{ab} & = & \mathcal{J}_{ab},\,\,\,\,\,\,\,Z_{ab}=\mathcal{Z}_{ab},\nonumber \\
P_{a} & = & \sqrt{\lambda}\left(\mathcal{J}_{5a}+\mathcal{J}_{6a}+\frac{1}{2}\mathcal{Z}_{5a}\right),
\end{eqnarray}
and assume a special condition as $\mathcal{Z}_{5a}=-\mathcal{Z}_{6a}$,
we get the minimal Maxwell algebra \citep{soroka2005,azcarraga2011}
as follows,
\begin{eqnarray}
\left[M_{ab},M_{cd}\right] & = & i\left(\eta_{ad}M_{bc}+\eta_{bc}M_{ad}-\eta_{ac}M_{bd}-\eta_{bd}M_{ac}\right),\nonumber \\
\left[M_{ab},P_{c}\right] & = & i\left(\eta_{bc}P_{a}-\eta_{ac}P_{b}\right),\nonumber \\
\left[P_{a},P_{b}\right] & = & i\lambda Z_{ab},\nonumber \\
\left[M_{ab},Z_{cd}\right] & = & i\left(\eta_{ad}Z_{bc}+\eta_{bc}Z_{ad}-\eta_{ac}Z_{bd}-\eta_{bd}Z_{ac}\right).
\end{eqnarray}
 Moreover, it is easy to see that these algebras are reduced to their
standard forms when taking $\mathcal{Z}_{AB}=0$.

\section{Gauging the maxwell-conformal algebra}

In the previous section, we have obtained the Maxwell extension of
the conformal algebra in Eq.(\ref{eq: mc algebra}). This may lead
to interesting results because we know that the conformal transformations
play an important role in various areas of physics. In particular,
the conformal transformations provide a useful framework to develop
generalized theories of gravity. In this context, we will establish
a gravitational gauge theory based on the Maxwell-conformal group.
So we start with $\mathfrak{mc}\left(1,3\right)$ algebra that acts
on a four-dimensional space-time. We will follow the similar methods
for constructing a gauge theory of gravity as in \citep{yang1954,utiyama1956,kibble1961,kaku1977,lord1985}
by using differential forms \citep{azcarraga2011,cebecioglu2014}.
Let us first define the Maxwell-conformal algebra-valued one-form
gauge field $\mathcal{A}\left(x\right)=\mathcal{A}^{A}\left(x\right)X_{A}$
as follows,

\begin{equation}
\mathcal{A}\left(x\right)=e^{a}P_{a}+c^{a}K_{a}+\chi D-\frac{1}{2}\omega^{ab}M_{ab}+B^{ab}Z_{ab}+r^{a}Z_{a}+\tilde{r}^{a}\tilde{Z}_{a}+rZ,\label{eq: gauge fields}
\end{equation}
where $\mathcal{A}^{A}\left(x\right)=\left\{ e^{a},c^{a},\chi,\omega^{ab},B^{ab},r^{a},\tilde{r}^{a},r\right\} $
are the gauge fields which correspond to the generators $X_{A}=\left\{ P_{a},K_{a},D,M_{ab},Z_{ab},Z_{a},\tilde{Z}_{a},Z\right\} $,
respectively. Besides, the unit dimension of gauge fields are considered
as $\left[e^{a}\right]=L$, $\left[c^{a}\right]=L^{-1}$, $\left[r^{a}\right]=L$,
$\left[\tilde{r}^{a}\right]=L^{-1}$ and the remaining gauge fields
are dimensionless. The variation of the gauge field $\mathcal{A}\left(x\right)$
under a gauge transformation can be found by using the following formula,

\begin{equation}
\delta\mathcal{A}=-d\zeta-i\left[\mathcal{A},\zeta\right],\label{eq: trans A}
\end{equation}
where $\zeta\left(x\right)$ represents a $\mathcal{MC}\left(1,3\right)$-valued
zero-form gauge generator which is defined as follows,

\begin{equation}
\zeta\left(x\right)=y^{a}P_{a}+n^{a}K_{a}+\rho D-\frac{1}{2}\tau^{ab}M_{ab}+\phi^{ab}Z_{ab}+s^{a}Z_{a}+\tilde{s}^{a}\tilde{Z}_{a}+sZ,\label{eq: gauge generators}
\end{equation}
where $y^{a}$$\left(x\right)$, $n^{a}\left(x\right)$, $\rho$$\left(x\right)$,
$\tau^{ab}\left(x\right)$ $\phi^{ab}$$\left(x\right)$, $s^{a}\left(x\right)$,
$\tilde{s}^{a}$$\left(x\right)$ and $s\left(x\right)$, are the
parameters of the corresponding generators. Thus, by the help of Eqs.(\ref{eq: mc algebra})
and (\ref{eq: trans A}), one can find the transformation law of gauge
fields as follows,

\begin{eqnarray}
\delta e^{a} & = & -dy^{a}-\omega_{\,\,b}^{a}y^{b}-\chi y^{a}+\rho e^{a}+\tau_{\,\,b}^{a}e^{b},\nonumber \\
\delta c^{a} & = & -dn^{a}-\omega_{\,\,b}^{a}n^{b}+\chi n^{a}-\rho c^{a}+\tau_{\,\,b}^{a}c^{b},\nonumber \\
\delta\chi & = & -d\rho-2y^{a}c_{a}+2n^{a}e_{a},\nonumber \\
\delta\omega^{ab} & = & -d\tau^{ab}-\omega_{\,\,\,c}^{[a}\tau^{c|b]}-2y^{[a}c^{b]}+2e^{[a}n^{b]},\nonumber \\
\delta B^{ab} & = & -d\phi^{ab}-\omega_{\,\,\,c}^{[a}\phi^{c|b]}+\frac{1}{2}y^{[a}c^{b]}+y^{[a}\tilde{r}^{b]}+n^{[a}r^{b]}-\frac{\lambda}{2}y^{[a}e^{b]}-\frac{1}{2\lambda}n^{[a}c^{b]},\nonumber \\
 &  & +\tau_{\,\,\,c}^{[a}B^{c|b]}-\frac{1}{2}e^{[a}n^{b]}-e^{[a}\tilde{s}^{b]}-c^{[a}s^{b]},\nonumber \\
\delta r^{a} & = & -ds^{a}-\omega_{\,\,b}^{a}s^{b}-\chi s^{a}+2B_{\,\,\,b}^{a}y^{b}-ry^{a}+\frac{1}{2\lambda}\chi n^{a},\nonumber \\
 &  & +\tau_{\,\,b}^{a}r^{b}+\rho r^{a}-2\phi_{\,\,\,b}^{a}e^{b}+se^{a}-\frac{1}{2\lambda}\rho c^{a},\nonumber \\
\delta\tilde{r}^{a} & = & -d\tilde{s}^{a}-\omega_{\,\,b}^{a}\tilde{s}^{b}+\chi\tilde{s}^{a}+2B_{\,\,\,b}^{a}n^{b}+rn^{a}-\frac{\lambda}{2}\chi y^{a},\nonumber \\
 &  & +\tau_{\,\,b}^{a}\tilde{r}^{b}-\rho\tilde{r}^{a}-2\phi_{\,\,\,b}^{a}c^{b}-sc^{a}+\frac{\lambda}{2}\rho e^{a},\nonumber \\
\delta r & = & -ds-y^{a}c_{a}-2y^{a}\tilde{r}_{a}+2n^{a}r_{a}+n^{a}e_{a}-2s^{a}c_{a}+2\tilde{s}^{a}e_{a}.\label{eq: trans gauge field}
\end{eqnarray}

The curvature two-forms of the associated gauge fields $\mathcal{F}\left(x\right)=\mathcal{F}^{A}\left(x\right)X_{A}$
are defined to be 

\begin{eqnarray}
\mathcal{F}\left(x\right) & = & R\left(P\right)^{a}P_{a}+R\left(K\right)^{a}K_{a}+R\left(D\right)D-\frac{1}{2}R\left(M\right)^{ab}M_{ab}\nonumber \\
 &  & +R\left(Z\right)^{ab}Z_{ab}+R\left(Z\right)^{a}Z_{a}+R\left(\tilde{Z}\right)^{a}\tilde{Z}_{a}+R\left(Z\right)Z,
\end{eqnarray}
where $R\left(P\right)^{a}$, $R\left(K\right)^{a}$, $R\left(D\right)$,
$R\left(M\right)^{ab}$ $R\left(Z\right)^{ab}$, $R\left(Z\right)^{a}$,
$R\left(\tilde{Z}\right)^{a}$ and $R\left(Z\right)$ represent the
curvatures which comes from the associated generators. In order to
find the explicit forms of these curvatures, we use the following
structure equation,

\begin{equation}
\mathcal{F}=d\mathcal{A}+\frac{i}{2}\left[\mathcal{A},\mathcal{A}\right],
\end{equation}
and taking account of the gauge fields in Eq.(\ref{eq: gauge fields}),
the group curvatures are found to be,

\begin{eqnarray}
R\left(P\right)^{a} & = & de^{a}+\omega_{\,\,b}^{a}\wedge e^{b}+\chi\wedge e^{a},\nonumber \\
R\left(K\right)^{a} & = & dc^{a}+\omega_{\,\,b}^{a}\wedge c^{b}-\chi\wedge c^{a},\nonumber \\
R\left(D\right) & = & d\chi-2e^{a}\wedge c_{a},\nonumber \\
R\left(M\right)^{ab} & = & R\left(\omega\right)^{ab}-2e^{[a}\wedge c^{b]},\nonumber \\
R\left(Z\right)^{ab} & = & dB^{ab}+\omega_{\,\,\,\,c}^{[a}\wedge B^{c|b]}-\frac{\lambda}{2}e^{a}\wedge e^{b}-\frac{1}{2\lambda}c^{a}\wedge c^{b}\nonumber \\
 &  & +\frac{1}{2}e^{[a}\wedge c^{b]}+e^{[a}\wedge\tilde{r}^{b]}+c^{[a}\wedge r^{b]},\nonumber \\
R\left(Z\right)^{a} & = & dr^{a}+\omega_{\,\,b}^{a}\wedge r^{b}+\chi\wedge r^{a}-2B_{\,\,b}^{a}\wedge e^{b}-e^{a}\wedge r-\frac{1}{2\lambda}\chi\wedge c^{a},\nonumber \\
R\left(\tilde{Z}\right)^{a} & = & d\tilde{r}^{a}+\omega_{\,\,b}^{a}\wedge\tilde{r}^{b}-\chi\wedge\tilde{r}^{a}-2B_{\,\,b}^{a}\wedge c^{b}+c^{a}\wedge r+\frac{\lambda}{2}\chi\wedge e^{a},\nonumber \\
R\left(Z\right) & = & dr-e^{a}\wedge c_{a}-2e^{a}\wedge\tilde{r}_{a}+2c^{a}\wedge r_{a},\label{eq: curv}
\end{eqnarray}
where

\begin{equation}
R\left(\omega\right)^{ab}=d\omega^{ab}+\omega_{\,\,c}^{a}\wedge\omega^{cb},\label{eq: lorentz curvature}
\end{equation}
denotes the usual Riemann curvature tensor. We note that the curvatures
$R\left(P\right)^{a}$, $R\left(K\right)^{a}$, $R\left(D\right)$
and $R\left(M\right)^{ab}$ have the same form as in the standard
conformal gravity. Moreover, the new curvature $R\left(Z\right)^{ab}$
contains $\lambda$-dependent term that recalls the contribution to
the Riemann curvature tensor as in $\left(A\right)dS$ gravity. This
is the well-known characterstics of the Maxwell-like algebras. The
transformation properties of the two-form curvatures under the infinitesimal
gauge transformation can be found by the following expression and
Eq.(\ref{eq: gauge generators}),

\begin{equation}
\delta\mathcal{F}=i\left[\zeta,\mathcal{F}\right],
\end{equation}
and thus one can obtain,

\begin{eqnarray}
\delta R\left(P\right)^{a} & = & \tau_{\,\,\,c}^{a}R\left(P\right)^{c}+\rho R\left(P\right)^{a}-R\left(M\right)_{\,\,\,c}^{a}y^{c}-y^{a}R\left(D\right),\nonumber \\
\delta R\left(K\right)^{a} & = & \tau_{\,\,\,c}^{a}R\left(K\right)^{c}-\rho R\left(K\right)^{a}-R\left(M\right)_{\,\,\,c}^{a}n^{c}+n^{a}R\left(D\right),\nonumber \\
\delta R\left(D\right) & = & -2y^{a}R\left(K\right)_{a}+2n^{a}R\left(P\right)_{a},\nonumber \\
\delta R\left(M\right)^{ab} & = & \tau_{\,\,\,\,c}^{[a}R\left(M\right)^{c|b]}-2y^{[a}R\left(K\right)^{b]}-2n^{[a}R\left(P\right)^{b]},\nonumber \\
\delta R\left(Z\right)^{ab} & = & \tau_{\,\,\,\,c}^{[a}R\left(Z\right)^{c|b]}-R\left(M\right)_{\,\,\,\,c}^{[a}\phi^{c|b]}+\frac{1}{2}n^{[a}R\left(P\right)^{b]}-\frac{1}{2\lambda}n^{[a}R\left(K\right)^{b]}\nonumber \\
 &  & -\frac{\lambda}{2}y^{[a}R\left(P\right)^{b]}+\frac{1}{2}y^{[a}R\left(K\right)^{b]}+\tilde{s}^{[a}R\left(P\right)^{b]}+s^{[a}R\left(K\right)^{b]}-R\left(\tilde{Z}\right)^{[a}y^{b]}-R\left(Z\right)^{[a}n^{b]},\nonumber \\
\delta R\left(Z\right)^{a} & = & \tau_{\,\,\,c}^{a}R\left(Z\right)^{c}+\rho R\left(Z\right)^{a}-2\phi_{\,\,\,c}^{a}R\left(P\right)^{c}-\frac{\lambda}{2}\rho R\left(K\right)^{a}-y^{a}R\left(Z\right)\nonumber \\
 &  & -R\left(M\right)_{\,\,\,c}^{a}s^{c}-R\left(D\right)s^{a}+2R\left(Z\right)_{\,\,\,c}^{a}y^{c}+\frac{\lambda}{2}R\left(D\right)n^{a}+R\left(P\right)^{a}s,\nonumber \\
\delta R\left(\tilde{Z}\right)^{a} & = & \tau_{\,\,\,c}^{a}R\left(\tilde{Z}\right)^{c}-\rho R\left(\tilde{Z}\right)^{a}-2\phi_{\,\,\,c}^{a}R\left(K\right)^{c}+\frac{1}{2\lambda}\rho R\left(P\right)^{a}+n^{a}R\left(Z\right)\nonumber \\
 &  & -R\left(M\right)_{\,\,\,c}^{a}\tilde{s}^{c}+R\left(D\right)\tilde{s}^{a}+2R\left(Z\right)_{\,\,\,c}^{a}n^{c}-\frac{1}{2\lambda}y^{a}R\left(D\right)-R\left(K\right)^{a}s,\nonumber \\
\delta R\left(Z\right) & = & -y^{a}R\left(K\right)_{a}-2y^{a}R\left(\tilde{Z}\right)_{a}+2n^{a}R\left(Z\right)_{a}+n^{a}R\left(P\right)_{a}+2\tilde{s}^{a}R\left(P\right)_{a}-2s^{a}R\left(K\right)_{a}.\label{eq: var curv}
\end{eqnarray}
Taking the exterior covariant derivative of the curvatures, we can
find the Bianchi identities as

\begin{eqnarray}
\mathcal{D}R\left(P\right)^{a} & = & R\left(\omega\right)_{\,\,\,c}^{a}\wedge e^{c}+d\chi\wedge e^{a},\nonumber \\
\mathcal{D}R\left(K\right)^{a} & = & R\left(\omega\right)_{\,\,\,c}^{a}\wedge c^{c}-d\chi\wedge c^{a},\nonumber \\
\mathcal{D}R\left(D\right) & = & -2R\left(P\right)^{a}\wedge c_{a}+2e^{a}\wedge R\left(K\right)_{a},\nonumber \\
\mathcal{D}R\left(M\right)^{ab} & = & -2R\left(P\right)^{[a}\wedge c^{b]}-2R\left(K\right)^{[a}\wedge e^{b]},\nonumber \\
\mathcal{D}R\left(Z\right)^{ab} & = & R\left(\omega\right)_{\,\,c}^{[a}\wedge B^{c|b]}-\frac{\lambda}{2}R\left(P\right)^{[a}\wedge e^{b]}-\frac{1}{2\lambda}R\left(K\right)^{[a}\wedge c^{b]}+\frac{1}{2}R\left(P\right)^{[a}\wedge c^{b]}+\frac{1}{2}R\left(K\right)^{[a}\wedge e^{b]}\nonumber \\
 &  & +\frac{1}{2}R\left(P\right)^{[a}\wedge\tilde{r}^{b]}+\frac{1}{2}\mathcal{D}\tilde{r}{}^{[a}\wedge e^{b]}+\frac{1}{2}R\left(K\right)^{[a}\wedge r^{b]}+\frac{1}{2}\mathcal{D}r{}^{[a}\wedge c^{b]},\nonumber \\
\mathcal{D}R\left(Z\right)^{a} & = & R\left(\omega\right)_{\,\,\,c}^{a}\wedge r^{c}+d\chi\wedge r^{a}-2\mathcal{D}B_{\,\,\,c}^{a}\wedge e^{c}+2B_{\,\,\,c}^{a}\wedge R\left(P\right)^{c}\nonumber \\
 &  & -R\left(P\right)^{a}\wedge r+e^{a}\wedge\mathcal{D}r-\frac{\lambda}{2}\mathcal{D}\chi\wedge c^{a}+\frac{\lambda}{2}\chi\wedge R\left(K\right)^{a},\nonumber \\
\mathcal{D}R\left(\tilde{Z}\right)^{a} & = & R\left(\omega\right)_{\,\,\,c}^{a}\wedge\tilde{r}^{c}-d\chi\wedge\tilde{r}^{a}-2\mathcal{D}B_{\,\,b}^{a}\wedge c^{b}+2B_{\,\,\,b}^{a}\wedge R\left(K\right)^{b}\nonumber \\
 &  & +R\left(K\right)^{a}\wedge r-c^{a}\wedge\mathcal{D}r+\frac{1}{2\lambda}\mathcal{D}\chi\wedge e^{a}-\frac{1}{2\lambda}\chi\wedge R\left(P\right)^{a},\nonumber \\
\mathcal{D}R\left(Z\right) & = & -R\left(P\right)^{a}\wedge c_{a}+e^{a}\wedge R\left(K\right)_{a}-2R\left(P\right)^{a}\wedge\tilde{r}_{a}+2e^{a}\wedge\mathcal{D}\tilde{r}_{a}+2R\left(K\right)^{a}\wedge r_{a}-2c^{a}\wedge\mathcal{D}r_{a},
\end{eqnarray}
where $\mathcal{D}$ is the Lorentz-Weyl covariant derivative which
is defined as $\mathcal{D}\Phi=\left[d+\omega+w\left(\Phi\right)\chi\right]\Phi$.
Here, $\omega$ is the spin connection and $w$ is the conformal weight
of the corresponding field.

\section{Gravitational action for $\mathcal{MC}\left(4\right)$}

We introduce our discussion with a summary of the previous treatments
of the conformal gravity. The conformal transformation of the metric
tensor $g_{\mu\nu}\left(x\right)$ can be given as follows,

\begin{equation}
g_{\mu\nu}\left(x\right)\rightarrow\tilde{g}{}_{\mu\nu}\left(x\right)=\Omega^{2}\left(x\right)g_{\mu\nu}\left(x\right),\label{eq: trans metric}
\end{equation}
where $\Omega\left(x\right)$ is an arbitrary, positive and dimensionless
space-time function which is known as the conformal factor. The transformations
in Eq.(\ref{eq: trans metric}) preserve the light-cone structure
and the angles between vectors \citep{Blagojevic2002}. This feature
restricts to construct an invariant action under conformal transformations.
For instance, the well-known Einstein-Hilbert (EH) action does not
satisfy the local conformal invariance. In the literature, there are
several methods to overcome this problem but two of these have attracted
much attention. The first one is to construct an action linear in
curvature scalar together with a Brans-Dicke like \citep{brans1961}
scalar compensating field $\phi\left(x\right)$ having conformal weight
($-1)$. In the context of this method, Deser \citep{deser1970} and
Dirac \citep{dirac1973} used a conformal invariant $\phi^{2}R$ term
rather than $R$ in the EH action to satisfy local conformal invariance,
\begin{equation}
S_{BD}=\int d^{4}x\sqrt{-g}\phi^{2}R.
\end{equation}
By the help of this approach, the generalized versions of Einstein's
field equations were obtained \citep{Freund1974,bramson1974,omote1974,kasuya1975,Bicknell1976}.
The second one is to use a curvature-quadratic action without a compensating
field in four dimensions as follows,

\begin{equation}
S_{W}=-\frac{\alpha}{4}\int d^{4}x\sqrt{-g}C_{\mu\nu\rho\sigma}C^{\mu\nu\rho\sigma},
\end{equation}
where $C_{\mu\nu\rho\sigma}$ is the Weyl tensor and $\alpha$ is
a coupling constant. This theory is also known as ``Weyl-squared''
theory. This type of action does not provide an exact relation with
Einstein's gravity but it leads to higher-order field equations \citep{wehner1999}.
Employing quadratic gravity where terms quadratic in the curvature
tensor, a Weyl like (super)gravity is obtained as a gauge theory of
the (super)conformal group in \citep{Crispim-Romao1977,kaku1977}.

In this section, we want to construct a conformal invariant action
which contains EH term together with additional contributions. For
this purpose, starting from the invariance principle, our research
focuses primarily on a free gravitational theory based on the gauge
theory of the Maxwell-conformal group. We initially define a shifted
curvature having zero conformal weight as,
\begin{equation}
R\left(J\right)^{ab}=R\left(M\right)^{ab}+\mu R\left(Z\right)^{ab},\label{eq: shifted curvature}
\end{equation}
where $\mu$ is an arbitrary dimensionless constant. Taking account
of the Eq.(\ref{eq: var curv}) and considering the constraints $R\left(P\right)^{a}=0$,
$R\left(K\right)^{a}=0$, $R\left(Z\right)^{a}=0$, $R\left(\tilde{Z}\right)^{a}=0$,
it is easy to see that the gauge transformation of the shifted curvature
can be found as $\delta R\left(J\right)^{ab}=\tau_{\,\,\,\,c}^{[a}R\left(J\right)^{c|b]}$.
By the help of this definition, we can write the Euler or Gauss-Bonnet
like topological gravitational action as follows,

\begin{eqnarray}
S & = & \frac{1}{2\kappa\mu\lambda}\int R\left(J\right)\wedge^{*}R\left(J\right)\nonumber \\
 & = & \frac{1}{4\kappa\mu\lambda}\int\epsilon_{abcd}R\left(J\right)^{ab}\wedge R\left(J\right)^{cd}\nonumber \\
 & = & \frac{1}{4\kappa\mu\lambda}\int\epsilon_{abcd}R\left(M\right)^{ab}\wedge R\left(M\right)^{cd}+2\mu\epsilon_{abcd}R\left(M\right)^{ab}\wedge R\left(Z\right)^{cd}+\mu^{2}\epsilon_{abcd}R\left(Z\right)^{ab}\wedge R\left(Z\right)^{cd},\label{eq: action}
\end{eqnarray}
where $\kappa=8\pi Gc^{-4}$ is Einstein's gravitational constant
and $^{*}$ denoting the Hodge star dual operator. If we take the
constant $\mu=4$ and ignoring the total derivative term, the action
becomes

\begin{eqnarray}
S & =- & \frac{1}{4\kappa}\int\epsilon_{abcd}R\left(\omega\right)^{ab}\wedge e^{c}\wedge e^{d}-\lambda\epsilon_{abcd}e^{a}\wedge e^{b}\wedge e^{c}\wedge e^{d}\nonumber \\
 &  & +\frac{1}{\lambda^{2}}\epsilon_{abcd}R\left(\omega\right)^{ab}\wedge c^{c}\wedge c^{d}-\frac{2}{\lambda}\epsilon_{abcd}R\left(\omega\right)^{ab}\wedge e^{[c}\wedge\tilde{r}^{d]}-\frac{2}{\lambda}\epsilon_{abcd}R\left(\omega\right)^{ab}\wedge c^{[c}\wedge r^{d]}\nonumber \\
 &  & -\frac{4}{\lambda}\epsilon_{abcd}\mathcal{D}B^{ab}\wedge\mathcal{D}B^{cd}+4\epsilon_{abcd}\mathcal{D}B^{ab}\wedge e^{c}\wedge e^{d}+\frac{4}{\lambda^{2}}\epsilon_{abcd}\mathcal{D}B^{ab}\wedge c^{c}\wedge c^{d}\nonumber \\
 &  & -\frac{8}{\lambda}\epsilon_{abcd}\mathcal{D}B^{ab}\wedge e^{[c}\wedge\tilde{r}^{d]}-\frac{8}{\lambda}\epsilon_{abcd}\mathcal{D}B^{ab}\wedge c^{[c}\wedge r^{d]}\nonumber \\
 &  & -\frac{2}{\lambda}\epsilon_{abcd}e^{a}\wedge e^{b}\wedge c^{c}\wedge c^{d}+4\epsilon_{abcd}e^{a}\wedge e^{b}\wedge e^{[c}\wedge\tilde{r}^{d]}+4\epsilon_{abcd}e^{a}\wedge e^{b}\wedge c^{[c}\wedge r^{d]}\nonumber \\
 &  & -\frac{1}{\lambda^{3}}\epsilon_{abcd}c^{a}\wedge c^{b}\wedge c^{c}\wedge c^{d}+\frac{4}{\lambda^{2}}\epsilon_{abcd}c^{a}\wedge c^{b}\wedge e^{[c}\wedge\tilde{r}^{d]}+\frac{4}{\lambda^{2}}\epsilon_{abcd}c^{a}\wedge c^{b}\wedge c^{[c}\wedge r^{d]}\nonumber \\
 &  & -\frac{4}{\lambda}\epsilon_{abcd}e^{[a}\wedge\tilde{r}^{b]}\wedge e^{[c}\wedge\tilde{r}^{d]}-\frac{8}{\lambda}\epsilon_{abcd}e^{[a}\wedge\tilde{r}^{b]}\wedge c^{[c}\wedge r^{d]}-\frac{4}{\lambda}\epsilon_{abcd}c^{[a}\wedge r^{b]}\wedge c^{[c}\wedge r^{d]}.\label{eq: action_expand}
\end{eqnarray}
The first two terms correspond the Einstein-Hilbert action together
with a cosmological term, and the remaining terms contain the higher-order
curvature term and mixed terms with the new extra fields $B^{ab}\left(x\right)$,
$r^{a}\left(x\right)$ and $\tilde{r}^{a}\left(x\right)$ coupled
to the spin connection, vielbein and special conformal gauge field.
Also, the cosmological constant appeared in the action as linearly
and inversely. Similar formulation was analyzed in \citep{Alexander2019}
as a dynamical structure of quantum cosmological constant. 

The resulting action is conformal invariant but not invariant under
the local transformations of $\mathcal{MC}\left(1,3\right)$ group.
This means that the Maxwell parts of $\mathcal{MC}\left(1,3\right)$
symmetry group were broken. In this construction, the Maxwell fields
$B^{ab}\left(x\right)$, $r^{a}\left(x\right)$ and $\tilde{r}^{a}\left(x\right)$
were employed to satisfy the local conformal invariance of the action.
Therefore, we obtain an action that provide a useful framework to
construct various gravitational theories because it includes additional
tensor, vector and scalar fields which come from the Maxwell symmetry.
For instance, if both tensors $R(P)^{a}$ and $R(K)^{a}$ are set
simultaneously equal to zero, then the constraints of the theory yield
that their corresponding gauge fields $e^{a}\left(x\right)$ and $c^{a}\left(x\right)$
are related by \citep{z=000142o=00015Bnik2017}

\begin{equation}
c^{a}\left(x\right):=\lambda\varphi\left(x\right)e^{a},\label{eq: constraint 1}
\end{equation}
where $\varphi\left(x\right)$ is a Brans-Dicke like dimensionless
scalar compensating field having the following transformation,
\begin{equation}
\delta\varphi\left(x\right)=-2\rho\left(x\right)\varphi\left(x\right).
\end{equation}
Naturally, we can expect another relationship between gauge parameters
as in Eq.(\ref{eq: constraint 1}),

\begin{equation}
n^{a}\left(x\right):=\lambda\varphi\left(x\right)y^{a},
\end{equation}
Using these relationships, the action in Eq.(\ref{eq: action_expand})
takes the following form,

\begin{eqnarray}
S & =- & \frac{1}{4\kappa}\int\left(1+\varphi^{2}\right)\epsilon_{abcd}R\left(\omega\right)^{ab}\wedge e^{c}\wedge e^{d}-\lambda\left(1+\varphi^{2}\right)^{2}\epsilon_{abcd}e^{a}\wedge e^{b}\wedge e^{c}\wedge e^{d}\nonumber \\
 &  & -\frac{2}{\lambda}\epsilon_{abcd}R\left(\omega\right)^{ab}\wedge e^{[c}\wedge\left(\tilde{r}^{d]}+\lambda\varphi r^{d]}\right)-\frac{4}{\lambda}\epsilon_{abcd}\mathcal{D}B^{ab}\wedge\mathcal{D}B^{cd}+4\left(1+\varphi^{2}\right)\epsilon_{abcd}\mathcal{D}B^{ab}\wedge e^{c}\wedge e^{d}\nonumber \\
 &  & -\frac{8}{\lambda}\epsilon_{abcd}\mathcal{D}B^{ab}\wedge e^{[c}\wedge\left(\tilde{r}^{d]}+\lambda\varphi r^{d]}\right)+4\left(1+\varphi^{2}\right)\epsilon_{abcd}e^{a}\wedge e^{b}\wedge e^{[c}\wedge\left(\tilde{r}^{d]}+\lambda\varphi r^{d]}\right)\nonumber \\
 &  & -\frac{16}{\lambda}\epsilon_{abcd}e^{a}\wedge e^{b}\wedge\left(\tilde{r}^{c}+\lambda\varphi r^{c}\right)\wedge\left(\tilde{r}^{d}+\lambda\varphi r^{d}\right).\label{eq: action_expand-1}
\end{eqnarray}
Similarly, if we impose the following relation,

\begin{equation}
\tilde{r}^{a}+\lambda\varphi r^{a}:=0,\label{eq: constraint 2}
\end{equation}
the action reduces to a generalized version of the Brans-Dicke theory
of gravity,

\begin{eqnarray}
S & =- & \frac{1}{4\kappa}\int\left(1+\varphi^{2}\right)\epsilon_{abcd}R\left(\omega\right)^{ab}\wedge e^{c}\wedge e^{d}-\lambda\left(1+\varphi\right)^{2}\epsilon_{abcd}e^{a}\wedge e^{b}\wedge e^{c}\wedge e^{d}\nonumber \\
 &  & -\frac{4}{\lambda}\epsilon_{abcd}\mathcal{D}B^{ab}\wedge\mathcal{D}B^{cd}+4\left(1+\varphi^{2}\right)\epsilon_{abcd}\mathcal{D}B^{ab}\wedge e^{c}\wedge e^{d}.\label{eq: BD gravity}
\end{eqnarray}

This action coincides with the result given by Chamseddine \citep{chamseddine2003}
when the Maxwell extension suppressed and is a generalization of the
action given in \citep{cebecioglu2014}. Now we are in a position
to consider the field equations of the theory which can be derived
from a variational action principle. Taking into account of the Eq.
(\ref{eq: trans gauge field}), the equations of motion can be found
by the variation of the action in Eq.(\ref{eq: action}) with respect
to the gauge fields $\omega^{ab}\left(x\right),$ $e^{a}\left(x\right),$$B^{ab}\left(x\right)$
and $\varphi\left(x\right)$ respectively,

\begin{eqnarray}
\mathcal{D}R\left(J\right)^{ab}+4B_{\,\,\,e}^{[a}\wedge R\left(J\right)^{e|b]} & = & 0,\\
\epsilon_{abcd}R\left(J\right)^{ab}\wedge e^{c} & = & 0,\label{eq: em e}\\
\mathcal{D}R\left(J\right)^{cd} & = & 0,\\
\epsilon_{abcd}R\left(J\right)^{ab}\wedge e^{c}\wedge e^{d} & = & 0.
\end{eqnarray}
Furthermore, one can show that all these equations of motion verify
each other. If we consider a special solution $R\left(M\right)^{ab}=-4R\left(Z\right)^{ab}$,
the equations of motion are satisfied. Making use of the shifted curvature
and Eq.(\ref{eq: em e}), we get the following equation,

\begin{equation}
R\left(J\right)_{\,\,b}^{a}-\frac{1}{2}\delta_{\,\,b}^{a}R\left(J\right)=0.
\end{equation}
Passing from the tangent indices to world space-time indices \citep{azcarraga2011,cebecioglu2015}
as,
\begin{equation}
R\left(J\right)^{\mu\nu}=e_{a}^{\,\,\mu}e_{b}^{\,\,\nu}R\left(J\right)^{ab}=\frac{1}{2}R\left(J\right)_{\,\,\,\rho\sigma}^{\mu\nu}dx^{\rho}\wedge dx^{\sigma},
\end{equation}
then considering Eq.(\ref{eq: curv}) and Eq.(\ref{eq: lorentz curvature}),
one gets the following field equation,

\begin{equation}
R\left(\omega\right)_{\,\,\rho}^{\mu}-\frac{1}{2}\delta_{\,\,\rho}^{\mu}R\left(\omega\right)+6\lambda\left(1+\varphi^{2}\right)\delta_{\rho}^{\mu}=T_{\,\,\rho}^{\mu}\left(B\right),\label{eq: EFE}
\end{equation}
where,

\begin{equation}
T_{\,\,\rho}^{\mu}\left(B\right)=-4\left(e_{\,\,a}^{\mu}e_{\,\,b}^{\nu}\mathcal{D}_{[\rho}B_{\nu]}^{ab}-\frac{1}{2}\delta_{\,\,\rho}^{\mu}e_{\,\,a}^{\sigma}e_{\,\,b}^{\nu}\mathcal{D}_{[\sigma}B_{\nu]}^{ab}\right),
\end{equation}
here the term $6\lambda\left(1+\varphi^{2}\right)$ corresponds a
generalized cosmological term and $T_{\,\,\rho}^{\mu}\left(B\right)$
represents the tensorial contribution of the Maxwell symmetry. Therefore,
we demonstrated that the new extended framework leads to the generalized
Einstein field equation together with a coordinate dependent cosmological
term plus additional energy-momentum tensors. Moreover, in the limit
of $B^{ab}\left(x\right)=0$ and $\varphi\left(x\right)=0$, we get
the well-known gravitational field equation with the cosmological
constant,

\begin{equation}
R\left(\omega\right)_{\,\,\rho}^{\mu}-\frac{1}{2}\delta_{\,\,\rho}^{\mu}R\left(\omega\right)+6\lambda\delta_{\rho}^{\mu}=0,\label{eq: efe2}
\end{equation}

\section{Conclusion}

In four-dimensional space-time, as we mentioned before, there are
several restrictions on the construction of the conformally invariant
Einstein Hilbert action. To overcome these restrictions, according
to the classical conformal gauge theory, it is required to introduce
either a scalar compensating field which has a certain conformal weight
\citep{brans1961,dirac1973,deser1970} or modify the related gauge
symmetry \citep{wheeler1991,wehner1999}. Furthermore, it is known
from the literature that the cosmological constant problem may require
an alternative approach to gravity. In the present paper, we wanted
to find a locally conformal invariant theory of gravity especially
for two reasons. The first, we know that the conformal invariance
is very significant field in high-energy physics. According to the
literature, it was thought that the gauge theory of conformal group
is an important study area to resolve the quantum theory gravity and
the unification problems \citep{flato1970,Alfaro1980,obukhov1982,frankin1985,shtanov1994,hooft2011}.
The second, to cope with the difficulties to establishing conformal
invariant gravity theories, we looked for to construct an alternative
way to obtain a gravitational action including the Einstein-Hilbert
term which is invariant under local conformal transformations by using
an extended gauge group. From these motivations, we have found a Maxwell
extension of the conformal group in Eq.(\ref{eq: mc algebra}). This
is done by considering the isomorphism between $\mathcal{SO}(2,4)$
and $\mathcal{C}(1,3)$. We then established a gauge theory of gravity
based on the Maxwell-conformal group. We found a local conformally
invariant gravitational action by breaking Maxwell part of related
symmetry group in Eq.(\ref{eq: action_expand}). The resulting gravitational
action contains the Einstein-Hilbert term together with a cosmological
constant and additional source terms. This action was constructed
by using the shifted curvature two-forms in Eq.(\ref{eq: shifted curvature})
which was constituted by the curvatures of the $\mathcal{MC}\left(1,3\right)$. 

In this theory, there are additional tensor, vector, and scalar fields
in comparison with the conformally invariant Brans-Dicke theory gravity.
These properties of the theory provide more general mathematical framework
to obtain different kind of gravity theories. Using these tools, in
a special condition given in Eq.(\ref{eq: constraint 1}) and Eq.(\ref{eq: constraint 2}),
this extended gravitational theory was reduced to a generalization
of Brans-Dicke theory of gravity \citep{O'Hanlon1973,maeda1989} in
Eq.(\ref{eq: BD gravity}). Subsequently, we obtained a new generalization
of Einstein's field equations which contain a dynamical cosmological
term and additional energy-momentum tensor Eq.(\ref{eq: EFE}). We
also obtained the standard formulation of Einstein's field equation
with a cosmological constant in the limit of $B^{ab}\left(x\right)=0$
and $\varphi\left(x\right)=0$ in Eq.(\ref{eq: efe2}).

In conclusion, we have presented a new conformally invariant geometric
structure for gravitational theory including a generalized cosmological
constant in the framework of the Maxwell-conformal algebra. Therefore
the gravitational theory constructed here may play a role in the resolution
of the mentioned problems. On the other hand, if one wants to construct
an invariant action under local $\mathcal{MC}\left(1,3\right)$ symmetry,
the Stelle-West like method \citep{kerrick1995,z=000142o=00015Bnik2017,stelle1980}
provides a useful background. According to this method, one can construct
an action in six-dimensional space and making dimensional reduction
to four-dimensional framework, we hope that it is possible to find
an invariant action under the local $\mathcal{MC}\left(1,3\right)$
transformations. We are working on this approach.

We also note that decomposition of higher dimensional algebras may
allow a powerful method to obtain various types of the Maxwell algebras.
In the spirit of this approach, analogously to the construction of
$\mathcal{MC}\left(1,3\right)$, our studies on this subject are in
progress.

\section*{Acknowledgments}

This study is supported by the Scientific and Technological Research
Council of Turkey (TÜB\.{I}TAK) Research project No. 118F364.

\end{document}